\documentclass[10pt, conference, compsocconf]{IEEEtran}
\ifCLASSINFOpdf
\else
\fi

\usepackage{graphicx}
\usepackage{url}
\usepackage{mdwlist} 
\usepackage{flushend}

\begin{document}
%
\title{Building an Expert System for Evaluation of Commercial Cloud Services}


\author{\IEEEauthorblockN{Zheng Li}
\IEEEauthorblockA{School of CS\\
NICTA and ANU\\
Canberra, Australia\\
Zheng.Li@nicta.com.au}
\and
\IEEEauthorblockN{Liam O'Brien}
\IEEEauthorblockA{School of CS\\
CECS, ANU\\
Canberra, Australia\\
liamob99@hotmail.com}
\and
\IEEEauthorblockN{Rainbow Cai}
\IEEEauthorblockA{Division of Information\\
Information Services, ANU\\
Canberra, Australia\\
Rainbow.Cai@anu.edu.au}
\and
\IEEEauthorblockN{He Zhang}
\IEEEauthorblockA{School of CSE\\
NICTA and UNSW\\
Sydney, Australia\\
He.Zhang@nicta.com.au}
}


%


\maketitle

\begin{abstract}
Commercial Cloud services have been increasingly supplied to customers in industry. To facilitate customers' decision makings like cost-benefit analysis or Cloud provider selection, evaluation of those Cloud services are becoming more and more crucial. However, compared with evaluation of traditional computing systems, more challenges will inevitably appear when evaluating rapidly-changing and user-uncontrollable commercial Cloud services. This paper proposes an expert system for Cloud evaluation that addresses emerging evaluation challenges in the context of Cloud Computing. Based on the knowledge and data accumulated by exploring the existing evaluation work, this expert system has been conceptually validated to be able to give suggestions and guidelines for implementing new evaluation experiments. As such, users can conveniently obtain evaluation experiences by using this expert system, which is essentially able to make existing efforts in Cloud services evaluation reusable and sustainable.

\end{abstract}


%
\IEEEpeerreviewmaketitle

\section{Introduction}
\label{I}

Since Cloud Computing has become increasingly accepted as one of the most promising computing paradigms in industry \cite{Buyya_2009}, providing Cloud services also becomes an emerging business. An increasing number of providers have started to supply commercial Cloud services with different terminologies, definitions, and goals \cite{Prodan_2009}. As such, evaluation of those Cloud services would be crucial for many purposes ranging from cost-benefit analysis for Cloud Computing adoption to decision making for Cloud provider selection. However, evaluation of commercial Cloud services is different to and more challenging than that of other computing systems. There are three main reasons for this:

\begin{itemize}
\renewcommand{\labelitemi}{$\bullet$}
\item
In contrast with traditional computing systems, the Cloud is relatively chaos \cite{Stokes_2011}. There is still a lack of standard definition of Cloud Computing, which inevitably leads to market hype and also skepticism and confusion \cite{Zhang_Cheng_2010}. As a result, it is hard to point out the range of Cloud Computing, and not to mention a specific guideline to evaluate different commercial Cloud services. Consequently, although we have already learned rich lessons from the evaluation of traditional computing systems \cite{Jain_1991,Obaidat_2010}, it is still necessary to accumulate evaluation experiences in the Cloud Computing domain.
\item
Evaluation results could be invalid soon after the evaluation and then not reusable. Cloud providers may continually upgrade their hardware and software infrastructures, and new commercial Cloud services may gradually enter the market. Hence, previous evaluation results can be quickly out of date as time goes by. For example, at the time of writing, Google is moving its App Engine service from CPU usage model to instance model \cite{Alesandre_2011}; Amazon is still acquiring additional sites for Cloud data center expansion \cite{Miller_2011}; while IBM just offered a public and commercial Cloud \cite{Harris_2011}. As a result, customers would have to continually re-design and repeat evaluation for employing commercial Cloud services.
\item
The back-ends (e.g. configurations of physical infrastructure) of commercial Cloud services are not controllable from the perspective of customers. Unlike consumer-owned computing systems, customers have little knowledge and control over the precise nature of Cloud services even in the \textquotedblleft locked down\textquotedblright { }environment \cite{Sobel_2008}. Evaluations in the context of public Cloud Computing are then inevitably more challenging than that for systems where the customer is in direct control of all aspects \cite{Stantchev_2009}. In fact, it is natural that the evaluation of uncontrollable systems would be more complex than that of controllable ones.
\end{itemize}

Therefore, particularly for commercial Cloud services, it is necessary to find a way to facilitate evaluation, and make existing evaluation efforts reusable and sustainable. This paper suggests an expert system for Cloud evaluation to address the aforementioned issues. This expert system concentrates on processes and experiences rather than results of Cloud services evaluation. When it comes to the general implementation process of Cloud services evaluation, we can roughly draw six common steps following the systematic approach to performance evaluation of computer systems \cite{Jain_1991}, as specified below and illustrated in Figure \ref{fig>1}:

\begin{figure}
\centering
\includegraphics[height=6.5cm]{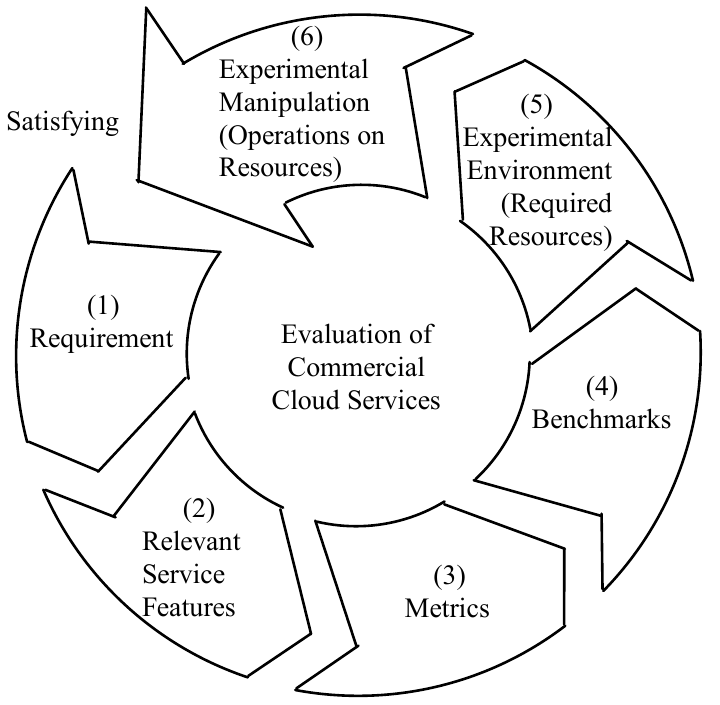}
\caption{\label{fig>1}General process of an evaluation implementation.}
\end{figure}

\begin{enumerate}
\renewcommand{\labelenumi}{\it{(\theenumi)}}
    \item	First of all, the requirement should be specified to clarify the evaluation purpose, which essentially drives the remaining steps of the evaluation implementation.
    \item	Based on the evaluation requirement, we can identify the relevant Cloud service features to be evaluated.
    \item	To measure the relevant service features, suitable metrics should be determined.
    \item	According to the determined metrics, we can employ corresponding benchmarks that may already exist or have to be developed.
    \item	Before implementing the evaluation experiment, the experimental environment should be constructed. The environment includes not only the Cloud resources to be evaluated but also assistant resources involved in the experiment.
    \item	Given all the aforementioned preparation, the evaluation experiment can be done with human intervention, which finally satisfies the evaluation requirement.
\end{enumerate}

Through decomposing and analyzing individual evaluation experiments following the six steps, we have collected and arranged data of detailed evaluation processes. Based on the primary evaluation data, general knowledge about evaluating commercial Cloud services can be abstracted and summarized. After manually constructing the \textit{Data/Knowledge Base}, we can design and implement an \textit{Inference Engine} to realize knowledge and data reasoning respectively. As such, given particular enquiries, the proposed expert system is not only able to supply common evaluation suggestions directly, but also able to introduce similar experimental practices to users for reference.

The remainder of this paper is organized as follows. Section \ref{III} specifies the establishments of \textit{Data Base}, \textit{Knowledge Base}, and \textit{Inference Engine} in this expert system. Section \ref{IV} employs three samples to show different application cases of this expert system, which also gives our current work a conceptual validation. Conclusions and some future work are discussed in Section \ref{V}.

\section{Establishment of this Expert System}
\label{III}

\begin{figure}
\centering
\includegraphics[height=4.5cm]{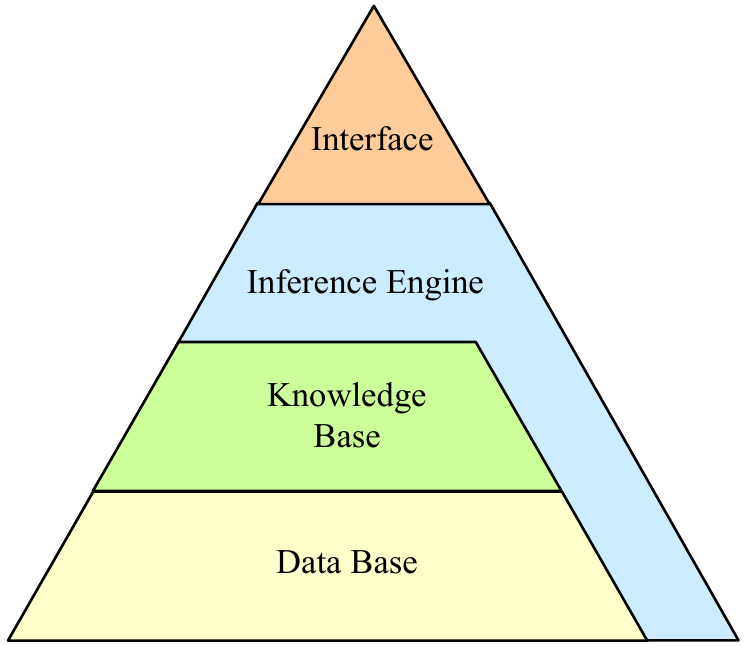}
\caption{\label{fig>3}Structure of this expert system.}
\end{figure}

Similar to general expert systems \cite{Jackson_1999,Kendal_2007}, the expert system proposed in this paper also comprises an \textit{Interface} with which users interact, an \textit{Inference Engine} that performs knowledge/data reasoning, and a \textit{Knowledge Base} that stores common and abstracted knowledge about evaluation of commercial Cloud services. However, we did not employ a specific knowledge acquisition module for building up the \textit{Knowledge Base} in this case. At the current stage, instead of obtaining knowledge by interviewing external experts, we extracted Cloud evaluation knowledge only from the collected data of published experimental studies. Moreover, for the convenience of acquiring experimental references, a \textit{Data Base} is maintained in this expert system to store initially-analyzed details of existing evaluation experiments. The complete structure of this expert system can be illustrated as shown in Figure \ref{fig>3}.

Considering the \textit{Interface} of this expert system can be designed at last in the future, this paper only specifies how we are realizing the \textit{Data/Knowledge Base} and \textit{Inference Engine}.

\subsection{Collecting Evaluation Practices (Data Base)}
\label{III-i}

To collect and initially analyze existing evaluation practices, we employed the systematic literature review (SLR) as the main approach. SLR is the methodology applied for Evidence-Based Software Engineering (EBSE) \cite{Dyba_2005}, and has been widely accepted as a standard and systematic way to investigate specific research questions by identifying, assessing, and analyzing published primary studies. According to the guidelines of SLR \cite{Kitchenham_2007}, an entire SLR instance mainly requires three stages, namely Preparation, Implementation, and Summarization. After adjusting some steps, here we list a rough SLR procedure suitable for this work:
\\

\textbf{\textit{Planning Review:}}
\begin{itemize}
\renewcommand{\labelitemi}{$\bullet$}
    \item	Justify the necessity of carrying out this SLR.
    \item	Identify research questions for this SLR.
    \item	Develop SLR protocol by defining search strategy, selection criteria, quality assessment standard, and data extraction schema for Conducting Review stage.
\end{itemize}

\textbf{\textit{Conducting Review:}}
\begin{itemize}
\renewcommand{\labelitemi}{$\bullet$}
    \item	Exhaustively search relevant primary studies in the literature.
    \item	Select relevant primary studies and assess their qualities for answering research questions.
    \item	Extract useful data from the selected primary studies.
    \item	Arrange and synthesize the initial results of our study into review notes.
\end{itemize}

\textbf{\textit{Reporting Review:}}
\begin{itemize}
\renewcommand{\labelitemi}{$\bullet$}
    \item	Analyze and interpret the initial results together with review notes into interpretation notes.
    \item	Finalize and polish the previous notes into an SLR report.
\end{itemize}

	\setcounter{footnote}{0}
Due to the limit of space, the detailed SLR process is not elaborated in this paper.\footnote{The complete SLR report can be found online: \url{https://docs.google.com/open?id=0B9KzcoAAmi43LV9IaEgtNnVUenVXSy1FWTJKSzRsdw}} With the pre-defined search strategy and rigorous selection criteria, we have identified 46 relevant primary studies covering six commercial Cloud providers from a set of popular digital publication databases (the studies are listed online for reference: \url{http://www.mendeley.com/groups/1104801/slr4cloud/papers/}). Note that this work focused only on the commercial Cloud services to make our effort closer to industry's needs. Moreover, this study paid attention to Infrastructure as a
Service (IaaS) and Platform as a Service (PaaS) without concerning
Software as a Service (SaaS). Since SaaS is not used to
further build individual business applications \cite{Binnig_Kossmann_2009}, various
SaaS implementations may comprise infinite and exclusive
functionalities to be evaluated, which could make our SLR out
of control even if adopting extremely strict selection/exclusion
criteria. After exhaustively identifying evaluation practices, independent evaluation experiments can be first isolated, and then be broken into atomic components by using the data extraction schema. For example, more than 500 evaluation metrics including duplications were finally extracted from the identified Cloud services evaluation studies. The summarized metrics in turn can help facilitate metric selection in future evaluation work \cite{Li_OBrien_2012b}. In other words, every single isolated experiment is finally represented as a set of elements, which essentially facilitates knowledge mining among different elements across different evaluation steps. Therefore, the Conducting Review stage in this SLR also indicates the procedure of establishing the \textit{Data Base} of the proposed expert system.

\subsection{Mining Evaluation Experiences (Knowledge Base)}
\label{III-ii}

Considering that this expert system gives evaluation suggestions according to users' inputs, the stored knowledge is suitable to be represented as rules each of which is composed of antecedent (to meet input) and consequent (to be output). To efficiently obtain rule-based knowledge from the \textit{Data Base}, we adopted association-rule mining to inspire the method of knowledge mining. By analogy, we can find that the rule-based knowledge used in this expert system has similar characteristics to association rules. As we know, different association rules can express different regularities that underlie a dataset \cite{Witten_2005}. In other words, association rules have freedom to use and predict any combination of items in a predefined set. As for the expert system, it is supposed that users can start from any particular evaluation step, or from any combination of detailed evaluation steps, to enquire about different evaluation experiences.

When mining association rules in a dataset, we use predefined \emph{coverage} to seek combinations of items that appear frequently, and then use \emph{accuracy} to extract suitable rules among each of the identified item combinations \cite{Witten_2005}. Note that we name an attribute-value pair as an item in the dataset. The \emph{coverage} of a particular item combination refers to the number of data instances comprising the combination, while the \emph{accuracy} of a candidate rule is the proportion of the correctly-predicted data instances to the applied data instances. 

Similarly, we followed the same rule-induction procedure to mine evaluation knowledge from the collected experiment data. However, compared with the quantitative and programmable process of rule mining, the knowledge mining in this case has to involve more human interventions. Although some evaluation step details (e.g. the Cloud service features) can be pre-standardized, most of the experiment data (e.g. requirement or experimental manipulation) would be not able to be specified within an exactly same schema from the beginning. As a result, we had to manually extract common knowledge through abstracting specific descriptions of the raw data. 

For the convenience of the discussion, we briefly demonstrate the process of knowledge mining only from two-item sets. Suppose the focus is now on the Scalability evaluation of Cloud services. We can initially list all the Scalability-related two-item combinations, gradually abstract their descriptions, and then rationally classify or unify them into fewer groups. At last, common knowledge can be sought within each of the combination groups. Here we only list four straightforward pieces of evaluation experience identified in our work.

\begin{itemize}
\renewcommand{\labelitemi}{$\bullet$}
    \item
\textbf{IF service feature = \textquotedblleft \textit{Scalability}\textquotedblright { }THEN experimental manipulation = \textquotedblleft \textit{varying Cloud resource with the same amount of workload}\textquotedblright} (extracted from the change of Cloud resource, without distinguishing that the resource was varied in terms of type or amount).
    \item
\textbf{IF service feature = \textquotedblleft \textit{Vertical Scalability}\textquotedblright { }THEN experimental manipulation = \textquotedblleft \textit{different types of Cloud resource}\textquotedblright} (extracted from the experiments each of which covers different types of service instances \cite{Lu_Jackson_2010,Rehr_2010}).
    \item
\textbf{IF service feature = \textquotedblleft \textit{Horizontal Scalability}\textquotedblright { }THEN experimental manipulation = \textquotedblleft \textit{different amount of Cloud resource}\textquotedblright} (extracted by focusing on the amount of the same type of Cloud resource, no matter the resource is CPU core \cite{Evangelinos_2008} or service instance \cite{Li_Humphrey_2010}).
    \item
\textbf{IF service feature = \textquotedblleft \textit{Scalability}\textquotedblright { }THEN metric = \textquotedblleft \textit{speedup over a baseline}\textquotedblright} (extracted by summarizing Scalability-related metrics like Pipeline Performance Speedup \cite{Li_Humphrey_2010}, Computation Speedup \cite{Rehr_2010}, and Throughput Speedup \cite{Lu_Jackson_2010}).
\end{itemize}

\subsection{Reasoning in Knowledge/Data Base (Inference Engine)}
\label{III-iii}
In this expert system, the \textit{Inference Engine} is associated with both \textit{Knowledge Base} and \textit{Data Base}. As such, the \textit{Inference Engine} can not only perform knowledge reasoning but also supply similar experimental cases.

\begin{enumerate}
\renewcommand{\labelenumi}{\it{\theenumi)}}
    \item	\emph{First-Order Rule Reasoning in Knowledge Base:}
\end{enumerate}

To facilitate the knowledge reasoning process, we proposed to enrich the representations of knowledge by bridging between some concepts or item combinations. The bridges can be viewed as common-sense knowledge that supplements the aforementioned, extracted knowledge. For example, we added two new rules \textbf{IF service feature = ``\textit{Vertical Scalability}" THEN service feature = ``\textit{Scalability}"} and \textbf{IF service feature = ``\textit{Horizontal Scalability}" THEN service feature = ``\textit{Scalability}"} to the four previous samples. In fact, the two new rules are always true although they are not generated by knowledge mining. Benefiting from the knowledge bridges, we can employ the algorithm of learning first-order rules \cite{Kotsiantis_2007} to conveniently reveal underlying rule-based knowledge that does not visibly exist in the \textit{Knowledge Base}. In this case, for instance, the expert system can give suggestions of evaluating Vertical Scalability by using a visible rule \textbf{IF service feature = ``\textit{Vertical Scalability}" THEN experimental environment = ``\textit{different types of Cloud resource}"}, and also resorting to the Scalability-related knowledge, as shown below.

\begin{itemize}
\renewcommand{\labelitemi}{$\bullet$}
    \item
Experimental manipulation = ``\textit{varying Cloud resource with the same amount of workload}".
    \item
Experimental environment = ``\textit{different types of Cloud resource}".
\item
Metric = ``\textit{speedup over a baseline}".
\end{itemize}

\begin{enumerate}
\renewcommand{\labelenumi}{\it{\theenumi)}}
\setcounter{enumi}{1}
    \item	\emph{Case Retrieving in Data Base:}
\end{enumerate}

\begin{figure}[!ht]
\centering
\includegraphics[height=6cm]{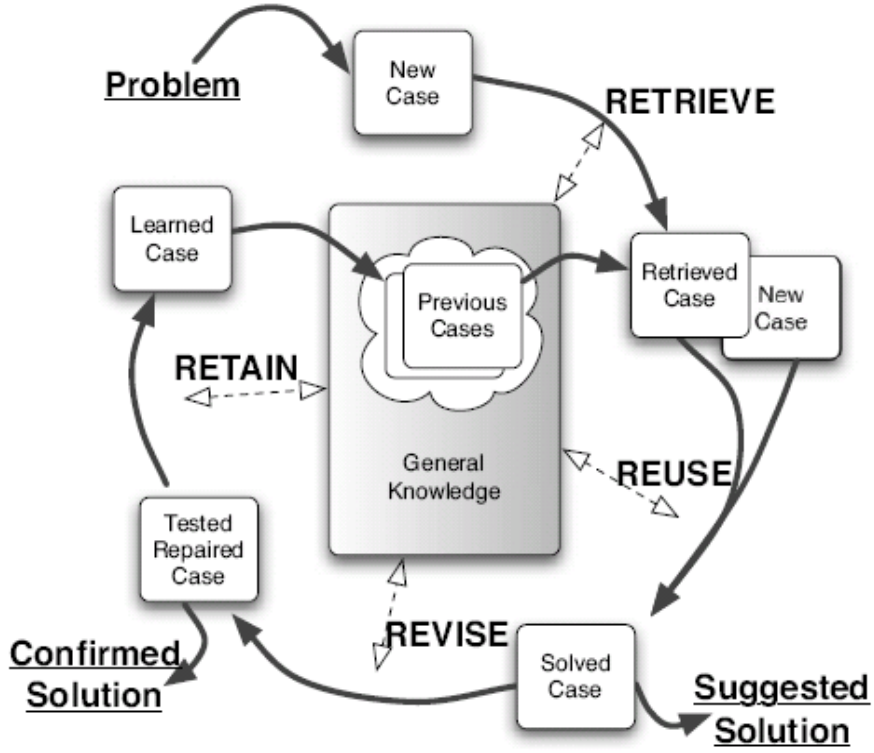}
\caption{\label{fig>4}General cyclical case-based reasoning process.}
\end{figure}

As mentioned previously, in addition to evaluation knowledge, this expert system is also able to retrieve experimental cases to users for analogy. In fact, as one of the basic human reasoning processes, analogy is used by almost every individual on a daily basis to solve new problems based upon past experiences. The process of analogy generally follows the procedure of the case-based reasoning (CBR), while one general CBR procedure comprises a four-stage cycle \cite{Aamodt_1994}, as shown in Figure \ref{fig>4}.

In the general cyclical CBR process, an initial problem is described as a
new case. Following the new case, we can RETRIEVE a case from the previous
cases. The retrieved case is combined with the new case through REUSE into a
solved case. The REVISE process is then used to test the solution for the new
case. Finally, useful experience is RETAINed for future reuse, and the dataset
of previous cases will be updated by a new learned case, or by modification of
some existing cases.

When it comes to case retrieving, the essential issue is how to identify rational and similar cases to the new one. We proposed three modes of case retrieving in this expert system, namely Precise Mode, Heuristic Mode, and Fuzzy Mode. As the name suggests, under the Precise Mode, the expert system identifies similar evaluation experiments exactly following users' enquiries. For example, suppose a user is interested in the evaluation experiments with respect to Horizontal Scalability of Cloud services, the expert system will only retrieve the experiment data with service feature = ``\textit{Horizontal Scalability}" 

\begin{figure*}[!t]
\centering
\includegraphics[height=4.3cm]{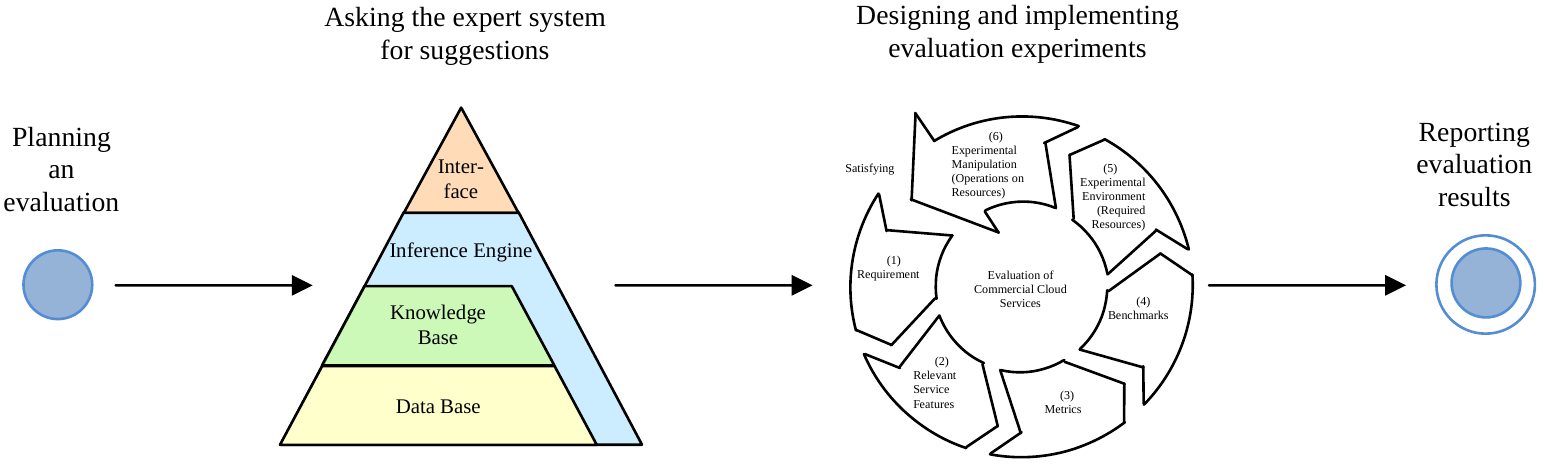}
\caption{\label{fig>5}Position of applying the expert system to Cloud services evaluation.}
\end{figure*}

In the worst case of Precise Mode, there would be no experiment record directly meeting a user's enquiry. The user can then try the Heuristic Mode. The Heuristic Mode relies on the knowledge reasoning process discussed previously. In detail, the expert system first explores the \textit{Knowledge Base} to identify the rules with antecedents meeting the user's enquiry, and then retrieves experiment data that include those rules' consequents. For example, when retrieving data with inquiry service feature = ``\textit{Horizontal Scalability}" under the Heuristic Mode, the expert system will list evaluation experiments having experimental environment = ``\textit{different amount of Cloud resource}" (according to the previous rule \textbf{IF service feature = ``\textit{Horizontal Scalability}" THEN experimental environment = ``\textit{different amount of Cloud resource}"}). In this case, suppose that even if the retrieved experiments focus on evaluating cost-benefit by using ``\textit{different amount of Cloud resource}", they can still be used to inspire the evaluation of Horizontal Scalability.

In the worst case of Heuristic Mode, the expert system could yet retrieve nothing due to lack of data, lack of knowledge, or invalid enquiry. Then, the data retrieving can be switched to the Fuzzy Mode. Ideally, the Fuzzy Mode relies on the uncertain reasoning \cite{Kendal_2007} in the \textit{Knowledge Base}. In the current stage, however, we only realize the Fuzzy Mode to allow the expert system to use sub-content of the enquiry information to explore for useful data. For example, suppose an invalid inquiry includes three elements: service feature = ``\textit{Horizontal Scalability}", experimental environment = ``\textit{different types of Cloud resource}", and metric = ``\textit{speedup over a baseline}". Under the Fuzzy Mode, the expert system first removes one of the inquiry elements, and then uses both Precise Mode and Heuristic Mode methods to identify similar experimental cases. In this sample, consequently, users can still achieve useful experimental cases after removing the inquiry element experimental environment = ``\textit{different types of Cloud resource}". Note that, since the case retrieving here is based on incomplete enquiry information, the Fuzzy Mode does not necessarily guarantee that all the retrieved experiments are valuable for users.

\section{Application of this Expert System}
\label{IV}

Ideally, this expert system is supposed to deal with enquiries about any component in evaluation experiments. For example, given a particular metric, we can ask the expert system for candidate benchmarks supplying the metric; or given particular experimental operations, we can ask the expert system for what evaluation requirement can be satisfied. Therefore, in general practices of Cloud services evaluation, the proposed expert system can be applied after planning an evaluation and before designing and implementing the evaluation, as illustrated in Figure \ref{fig>5}.

In the current stage of our work, we constrain the enquiry condition as: given a particular Cloud service feature, we ask the expert system for suitable evaluation scenarios (the combination of suitable experimental environment and experimental operations) and evaluation metrics (also relevant benchmarks). Here we use three real examples to show the possible application cases of this expert system. The three cases can meanwhile be viewed as a conceptual validation of our current work. Note that, to highlight the application flow, the expert system's working mechanism is simplified without elaborating the data/knowledge reasoning procedures.

\subsection{Application Case 1: Asking for evaluation suggestions before performing analytic modeling.}

\begin{figure*}
\centering
\includegraphics{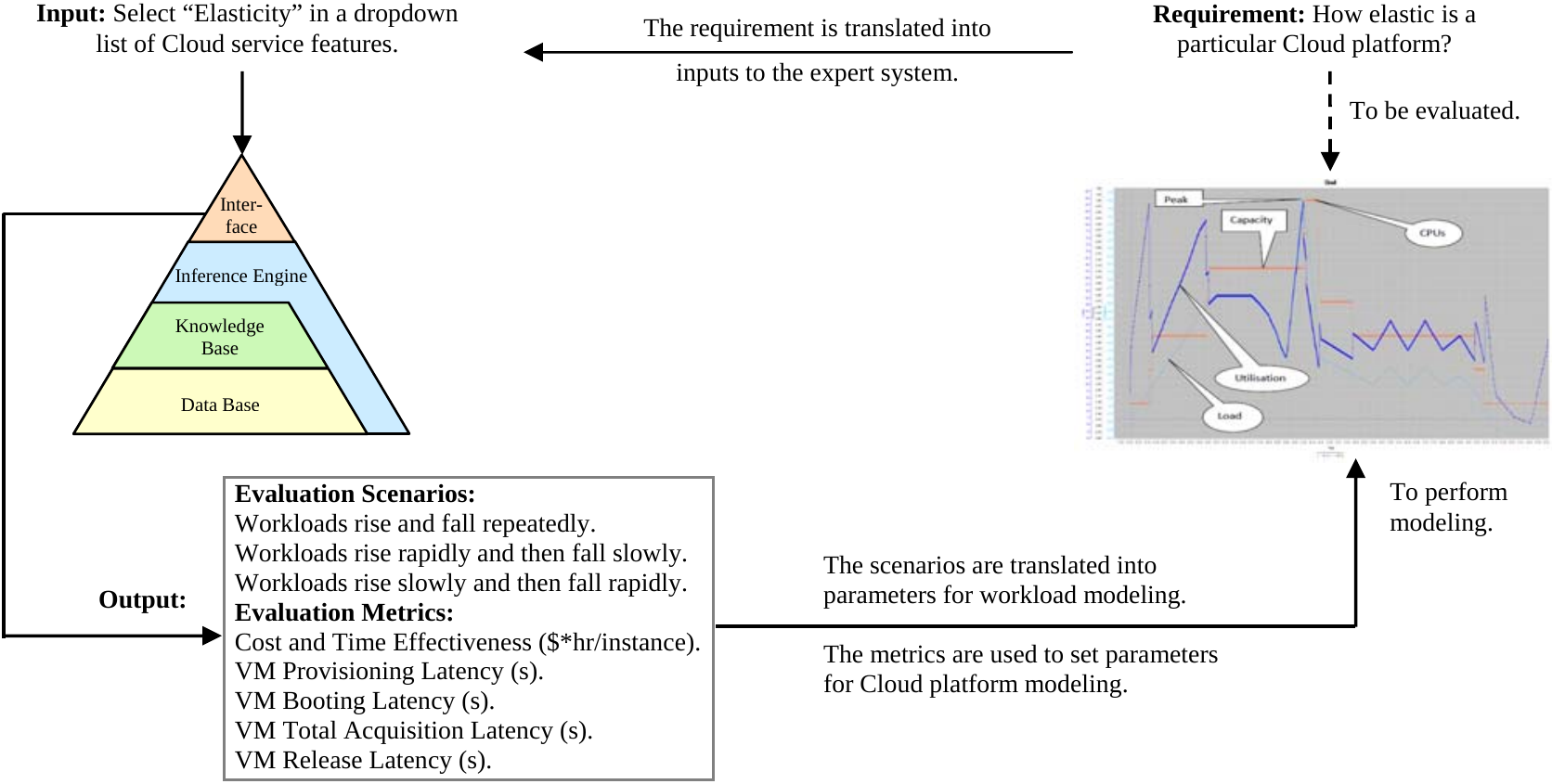}
\caption{\label{fig>6}Asking for evaluation suggestions before performing analytic modeling.}
\end{figure*}

Analytic modeling is a relatively light-weight evaluation technique, which employs approximate calculations to supply quick and rough analysis \cite{Obaidat_2010}. Suppose we decide to adopt analytic modeling to satisfy the requirement about how elastic a particular Cloud platform is \cite{Brebner_2011}. According to the keywords in the requirement, we can find that the concerned Cloud service feature is Elasticity in this case. As such, the requirement is manually translated into ``Elasticity" as the input to the expert system. The output includes scenarios and metrics for evaluating Elasticity, as demonstrated in Figure \ref{fig>6}. The evaluation scenarios and metrics can be further translated into design parameters and variables \cite{Obaidat_2010} by evaluator while performing the modeling work. The suggested metrics like \textit{VM Boosting Latency} are used to model the Cloud platform, while the suggested scenarios like \textit{Workloads rise and fall repeatedly} are used to model the Cloud-hosted workloads. The complete process of this application case is shown in Figure \ref{fig>6}.

\subsection{Application Case 2: Asking for evaluation suggestions before performing real measurement.}

Real measurement is a relatively effort-intensive evaluation technique, which implements experiments on prototypes or actual computing systems to conduct more accurate analysis \cite{Obaidat_2010}. In this case, suppose we plan to measure how variable the real Cloud service performance is \cite{Iosup_Yigitbasi_2010}. Similarly, the requirement here can be translated into ``Variability" as the input to the expert system, and the output suggests scenarios and metrics for evaluating Variability (cf. Figure \ref{fig>7}). Unlike the previous case, however, the suggested scenarios like \textit{Repeat experiment at different time} are used to prepare and perform experimental environment/operations, while the suggested metrics like \textit{Standard Deviation with Average Value} are used to measure and display the experimental results. The application flow of this case is shown in Figure \ref{fig>7}.

\subsection{Application Case 3: Asking for evaluation suggestions for further decision making about usage strategy of Cloud resources.}

This application case is essentially an extension of previous two cases. In fact, suggesting usage strategy of Cloud resources is out of the scope of applying this expert system. This expert system provides suggestions only for Cloud services evaluation, without making any decision based on the evaluation result. Nevertheless, since evaluation is the prerequisite of further decision making, this expert system can still be helpful in this case. Suppose there is an evaluation requirement about choosing ``alternative architecture for transaction processing in the Cloud" \cite{Kossmann_Kraska_2010}. This requirement is a typical decision making about alternative strategies of utilizing Cloud resources. Given particular Cloud service features concerned in the predefined architectures, the expert system supplies suggestions of evaluating those service features; the evaluation suggestions can then be employed in each experiment for each of the architectures respectively; the architectures are finally judged through contrasting the evaluation results. In other words, this application case normally comprises a set of sibling experiments with the same evaluation suggestions, and the application flow is similar to Figure \ref{fig>7}. As for the sample \cite{Kossmann_Kraska_2010}, this expert system will give suggestions of evaluation scenarios and metrics for the service features ``Storage" and ``Cost".\\

Overall, this section has demonstrated three typical cases of applying the proposed expert system. To better distinguish between these application cases, here we highlight two points:

\begin{figure*}
\centering
\includegraphics{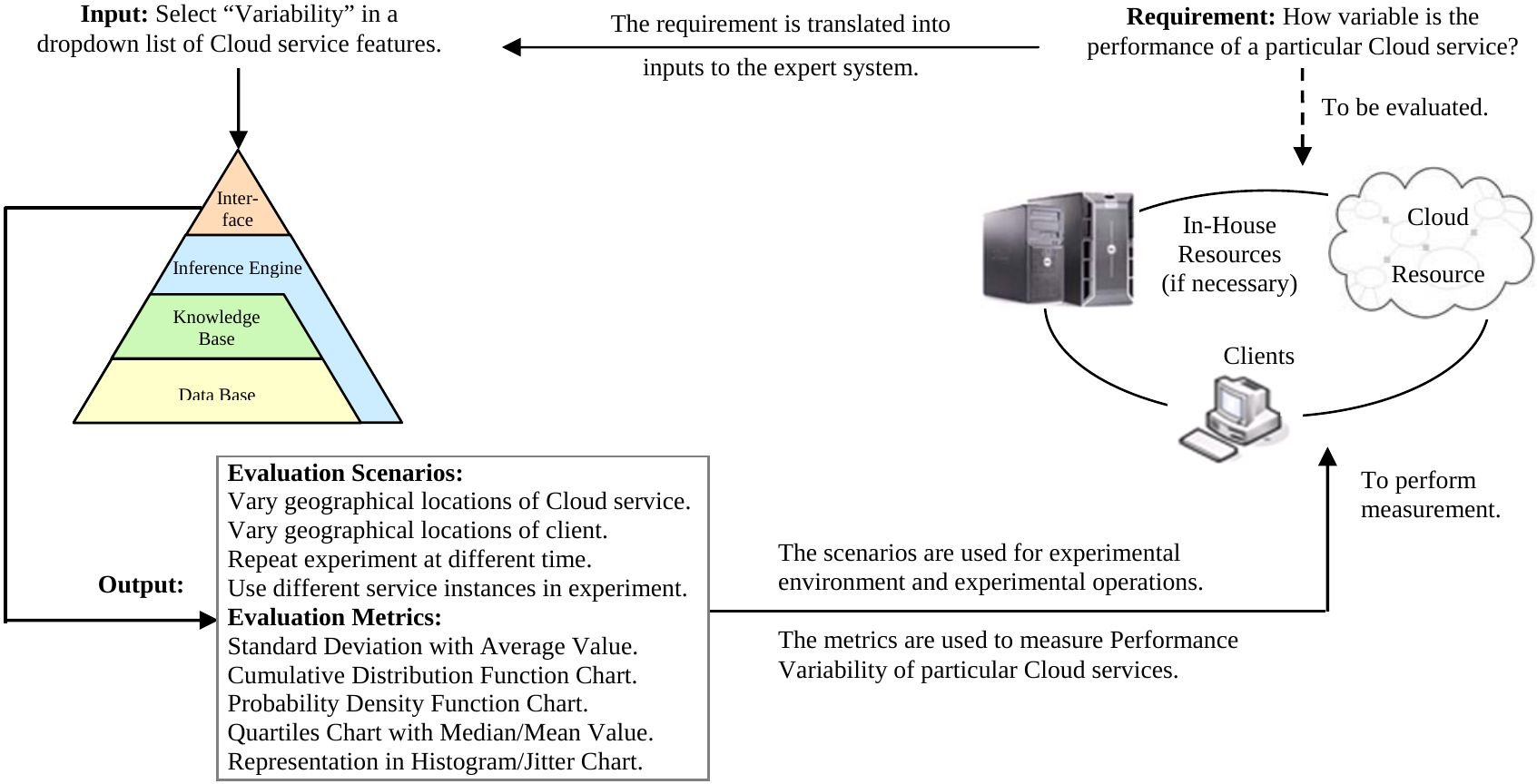}
\caption{\label{fig>7}Asking for evaluation suggestions before performing real measurement.}
\end{figure*}

\begin{itemize}
\renewcommand{\labelitemi}{$\bullet$}
    \item
\textit{\textbf{Direct vs. Indirect Help from the Expert System.}} When it comes to the application context of this expert system, Cases 1 and 2 are in the pure evaluation context, whereas Case 3 is in the decision making context. As previously mentioned, this expert system can facilitate but not directly suggest usage strategies of Cloud resources. Therefore, the expert system directly helps satisfy the evaluation requirement in the first two cases, while it indirectly helps satisfy the evaluation requirement in the third case.
    \item
\textit{\textbf{Different Usage Purpose and Sequence of the Evaluation Suggestions.}} In Application Case 1, the suggestion of evaluation metrics is used before the analytic modeling work. In fact, the suggested metrics are used to design the indicator tradeoffs of the simulated Cloud environment. On the contrary, in Application Case 2, the suggestion of evaluation metrics is used to measure the Cloud service indicators and display the result after real experiments.
\end{itemize}

\section{Conclusions and Future Work}
\label{V}

Along with the booming Cloud Computing in industry, various commercial providers have started offering Cloud services. Thus, it is normal and vital to implement evaluations when deciding whether or not to employ a particular Cloud service, or choosing among several candidate Cloud services. However, given the rapidly-changing and customer-uncontrollable conditions, evaluation of commercial Cloud services are inevitably more challenging than that of traditional computing systems. To facilitate evaluation work in the context of Cloud Computing in industry, we proposed to accumulate existing evaluation knowledge, and to establish an expert system for Cloud services evaluation to make evaluation experiences conveniently reusable and sustainable. Note that the proposed expert system does NOT work like an automated evaluation tool or benchmark involved in evaluation implementations, but gives evaluation suggestions or guidelines according to users' enquiries. 

The most significant contribution of this work is to help practitioners implement Cloud services evaluation along a systematic way. In fact, it is impossible to require everyone, especially Cloud customers, to be equipped with rich knowledge and expertise on Cloud services evaluation. We can find that the current evaluation work is relatively ad hoc in the Cloud computing field. For example, evaluation techniques and benchmarks are selected randomly. Based on the accumulated evaluation experiences, however, this proposed expert system can intelligently supply rational and comprehensive consultation to future evaluation practices, which has been conceptually validated by using three real application cases.

This paper roughly introduces the structure and components of this expert system, and mainly specifies the study methodology we are following. The methodology then reveals and guides our current and future work, such as using the SLR to collect and analyze existing evaluation practices, following the procedure of association-rule mining to extract evaluation knowledge, building the \textit{Inference Engine} to conduct knowledge reasoning and data retrieving, and patching a well-designed \textit{Interface} to complete the expert system. The prototypes of different function parts will be gradually developed and integrated into an online system.\footnote{The current version can be viewed as a knowledge lookup system: \url{http://sites.google.com/site/expertsystemforcloudevaluation/}} Finally, a data maintenance system will be also built up online to collect feedback, update data, and keep all the data versions with time stamps. The \textit{Data/Knowledge Base} of the expert system can eventually be updated regularly through the maintenance system.

\section*{Acknowledgment}

This project is supported by the Commonwealth of Australia under the Australia-China Science and Research Fund.

NICTA is funded by the Australian Government as represented by the Department of Broadband, Communications and the Digital Economy and the Australian Research Council through the ICT Centre of Excellence program.



%

\end{document}